# Comparing a Few Qubit Systems for Superconducting Hardware Compatibility and Circuit Design Sensitivity in Qiskit

Hillol Biswas,

Department of Electrical and Computer Engineering,

Democritus University of Thrace,

Xanthi, GREECE

***Abstract*—The development of complex circuits for practical applications in the current quantum computing ecosystem is based on basic primitives such as Bell states, which provide superposition, entanglement, and coherence. The range of domain-specific quantum applications has been greatly expanded by the availability of simulators and platforms such as IBM Quantum, which are supported by Qiskit. However, disparities between ideal simulator outputs and actual quantum processing unit (QPU) executions in the Noisy Intermediate-Scale Quantum (NISQ) era require the application of quantum error mitigation techniques. Limitations arise from hardware constraints in superconducting qubit systems and from the limited resources of classical simulators as quantum circuits grow. Quantum decoherence, which lowers gate fidelity and builds up at the circuit level with increasing depth, is specifically caused by material-induced flaws and interfaces. This creates a clear connection between circuit reliability, device performance, and material attributes. To address this, the current work uses both simulation and actual hardware on the IBM Sherbrooke 127-qubit processor to study three basic circuit classes over 4 to 10 qubits: the Quantum Fourier transform, the Greenberger–Horne–Zeilinger state, and the W state. The study examines trade-offs between circuit complexity, noise robustness, and resource utilization by contrasting simulator and QPU results. The results imply that circuit fidelity can serve as an indirect probe of material-limited noise, opening the door to a framework for designing quantum circuits that accounts for both hardware and materials to achieve scalable quantum advantage.**

## I. INTRODUCTION

The conceptual underpinnings of quantum mechanics have gradually evolved into useful frameworks for quantum computation since the experimental validation of nonlocal correlations in Bell test experiments, which challenged the idea of local hidden variables [1]. The foundation for representing and modifying quantum information in a structured computational form was laid by early advances, such as the creation of quantum circuits, including the Bell state [2].

Quantum devices, which use various physical systems to realize qubits, are at the heart of quantum computing. For example, a spin-1/2 particle's spin states can be used to encode computational basis states; the +1/2 and −1/2 spin states are mapped to $|0\rangle$ and $|1\rangle$ state, respectively. The adaptability of physical realizations for quantum information processing is further demonstrated by the fact that photon polarization, atomic or quantum-dot energy levels, and other quantum degrees of freedom can serve as qubit implementations.

Superconducting qubits have become a prominent architecture among these platforms, creating a new paradigm for scalable quantum computing. Specifically, fixed-frequency qubits, whose characteristics are specified during manufacture, are used in IBM Quantum systems. In these systems, the control qubit is driven at the target qubit's resonance frequency using the cross-resonance (CR) gate, a native two-qubit entangling operation [3]. Experimental demonstrations reveal gate fidelities of about 81% and the creation of entangled states with concurrence values up to 0.88 [4]. This all-microwave technique allows customizable interactions without the need for extra sub-circuitry. These developments highlight how superconducting platforms can be used to create quantum circuits.

At this crucial point in the research, quantum systems have grown large enough to solve computational problems beyond the scope of traditional brute-force techniques. In this regard, hybrid computational paradigms that strongly combine quantum and classical resources have been proposed, such as quantum-centric supercomputing (QCSC). To enable algorithmic advancements that neither classical nor quantum systems could achieve on their own, these frameworks seek to leverage the complementary strengths of both paradigms [5].

The quantum transpilation process is an essential part of connecting high-level quantum algorithms with physical hardware. The transpiler in Qiskit is a circuit-to-circuit rewriting tool that converts abstract quantum circuits into formats that are compatible with the instruction set architecture (ISA) of a target device. This approach is intrinsically dependent on the underlying hardware limits and requires several optimization steps. The executable instruction space for superconducting qubit systems is defined by the native gate set, which usually consists of operations like CNOT, √X, and parameterized rotations like RZ(θ) [6].

Additionally, Qiskit's Target abstraction, which simulates the capabilities and limitations of specific quantum devices, including gate availability, connectivity, timing, and error characteristics, enables hardware-aware circuit adaptation. Circuits can be optimized for a variety of platforms, including superconducting and trapped-ion systems, thanks to our retargetable transpilation framework, which guarantees effective execution across different hardware architectures [6].

The scalability of this technology is demonstrated by recent developments in superconducting quantum processors, such as the IBM Sherbrooke 127-qubit processor and the IBM Brisbane 127-qubit processor based on the Eagle r3 architecture. These processors are available on the IBM Quantum platform and feature single-qubit operations, including Rx, SX, and X gates, as well as an echoed cross-resonance (ECR) gate set [7], [8]. These systems enable the systematic study of circuit behavior under realistic noise and architectural constraints, providing a useful testbed for the implementation and analysis of quantum circuits on actual hardware.

This work is organized into the following four sections: Section II – Quantum Circuits, Qubit Devices and Superconducting Qubits, which describe the three chosen circuits, viz. QFT, GHZ, and W-state, Section III. Approach - Circuits in the simulator and the QPU are described in Section IV. Outcomes and Section V, Conclusion, wrap up the paper.

# II. Quantum Circuits, Qubit Devices and Superconducting Qubits

The theoretical groundwork for quantum computation was established by Richard Feynman's postulation of a quantum computer [9], which was followed over time by the creation of groundbreaking quantum algorithms such as the Deutsch algorithm [10], Bernstein–Vazirani algorithm [11], Deutsch–Jozsa algorithm [12], Shor Algorithm [13], Grover algorithm [14], and Simon’s algorithm [15]. These developments demonstrated the potential for exponential speedups over classical methods, spurring the investigation of practical physical platforms for implementing quantum systems.

Simultaneously, considerable efforts have been made to find appropriate material systems and device designs that can realize qubits. An important first step in this area was the exploration of superconducting platforms, especially Superconducting Quantum Interference Devices (SQUIDs), as well as superconducting qubits, such as charge qubits and Josephson-junction-based devices. Simultaneously, advances in molecular magnetism demonstrated that, starting in the 1990s, certain d-block metal-based polynuclear molecular clusters exhibit magnetic hysteresis at low temperatures, comparable to that of bulk magnetic materials. This finding paved the way for molecular quantum electronics. The 2000s saw the emergence of a second generation of molecular nanomagnets, mostly based on mononuclear complexes containing a single magnetic ion, usually from the f-block. Due to their potential for quantum information processing, including implementations aligned with algorithms such as Grover's search, these systems—especially those based on lanthanides—have attracted significant attention [16].

Because of their scalability and compatibility with current fabrication methods, superconducting qubit devices have gradually gained significance across a variety of platforms. The Josephson junction, in which a nanoscale superconducting electrode connected to a reservoir creates a single-Cooper-pair box and thereby realizes an artificial two-level electronic system, is a crucial enabling component [17]. These systems are good candidates for large-scale integration because they can be fabricated using traditional electron-beam lithography.

These systems often use qubits as micrometer-sized superconducting loops, broken into three or four Josephson junctions. The qubit states are defined by persistent currents in opposite directions. Coherent quantum superpositions are enabled by the controlled manipulation of these states via pulsed microwave modulation of the magnetic flux. Moreover, controlled magnetic flux transfer mediated by persistent currents, usually through superconducting coupling devices or flux transporters, facilitates entanglement between qubits [18].

Because of the underlying Josephson effect and the durability of superconductivity, superconducting qubit performance has increased by several orders of magnitude in recent years. Crucially, these systems have not yet reached basic physical limitations, indicating ongoing scaling. However, major architectural obstacles must be overcome to realize large-scale, fault-tolerant quantum processors. Physicists must specifically create and oversee intricate, actively managed systems that preserve coherence despite being intrinsically dissipative to conduct quantum error correction [19].

From an architectural perspective, expanded quantum bus designs have been used in industry initiatives, most notably by IBM, to show the scalability of superconducting qubits. In these arrangements, each qubit is coupled to a separate resonator for independent readout and control, and single-junction transmon qubits are connected via coplanar waveguide (CPW) resonators that serve as quantum buses [20]. This modular architecture enables multi-qubit systems to be integrated and scaled effectively.

The performance and scalability of superconducting quantum systems have improved with each generation of quantum processors IBM Quantum has developed. Transmon qubits are primarily used in contemporary devices because of their improved coherence and reduced sensitivity to charge noise, making them ideal for practical quantum computation [21].

Moreover, the development of fundamental quantum circuits and states, such as the quantum Fourier transform (QFT), Greenberger–Horne–Zeilinger (GHZ) states, and W states, has closely followed the progress of quantum computing hardware. These structures serve as the foundation for numerous quantum algorithms and protocols and are frequently used on modern quantum processing units, including IBM QPUs, which helps to close the gap between theoretical concepts and practical quantum applications.

This section is further split into five subsections: A. QFT Circuit, B. GHZ circuit, C. W state circuit, D. Many-qubit systems of circuits, and E. Error correction, mitigation, and work contribution, as given below.

## A. QFT circuit

Since the introduction of QFT in [22], Eq. 1 as given below, it is considered in many quantum computing applications, as in Eq. 1:

$$\frac{1}{\sqrt{q}}\sum_{c=0}^{q-1} e^{2\pi iac/q}|c\rangle \quad (1)$$

The QFT equation, which transforms a computational basis state into a superposition with phase factors encoding frequency-domain information, is shown above.

Numerous optimized QFT implementations have been put out to increase practical realisability. One popular method is the semiclassical (measurement-based) QFT, which substitutes sequential single-qubit measurements with classical feedforward for controlled rotation gates. This technique efficiently reduces the quantum circuit depth while increasing reliance on classical post-processing by applying conditional phase corrections based on previous measurement results [23].

Approximate QFT methods that exclude controlled-phase gates corresponding to extremely small rotation angles require additional optimization. This truncation greatly reduces circuit complexity while introducing a bounded error ε. For large qubit numbers, this results in exponential savings as the computational cost drops from $O(n^2)$ to $O(n \log n)$ [24]. For near-term quantum systems with limited gate fidelity and coherence duration, these approximations are especially useful.

The Quantum Computation Simulator (QCS) in MATLAB was used to develop a QFT for a 2-qubit system as part of early experimental and simulation-based validations [25]. QFT has been widely used in Qiskit for a variety of applications, such as n-qubit shift operations and composite algorithmic routines, as quantum software frameworks have advanced [26], [27].

Effectively mapping QFT circuits onto actual hardware architectures remains a significant challenge. Linear-depth QFT implementations for 5-qubit systems have been studied on a variety of platforms, including IBM's HeavyHex, Google's Sycamore, and two-dimensional grid designs. Furthermore, using conventional structural templates, formal approaches to program synthesis have been developed to find optimal QFT mappings under various architectural constraints [28].

Dynamic quantum circuit integration has achieved notable performance gains in recent QFT implementations. Specifically, IBM superconducting quantum hardware tests have surpassed previous benchmarks across quantum computing platforms, with certified process fidelities exceeding 50% for systems up to 16 qubits and above 1% for systems up to 37 qubits [29].

QFT provides benefits in resource optimization in addition to algorithmic efficiency. Phase rotation gates can directly include classical information, allowing QFT-based arithmetic operations to be carried out between quantum states and classical inputs without the need for extra qubits. Additionally, QFT-based quantum-classical comparator designs have been developed and expanded to facilitate quantum integer comparisons and modular arithmetic. These implementations are particularly resource-efficient, using only one auxiliary qubit, making them ideal for contexts with limited quantum hardware [30].

## B. GHZ state circuit

When extended to quantum systems with at least three particles, it is demonstrated that the tenets of the Einstein-Podolsky-Rosen paper are incoherent. The proof shows that, even in the case of perfect correlations, the EPR program defies quantum mechanics [31] and subsequent papers [32], [33]. For an n-qubit system, the (2) is as written as:

$$\frac{1}{\sqrt{n}}\left( |0\rangle^{\otimes n} + |1\rangle^{\otimes n}\right) \quad (2)$$

In solid-state circuit quantum electrodynamics (QED) systems, quantum control methods have been developed to deterministically produce and stabilize three-qubit GHZ states. These methods enable the controlled generation of GHZ states on superconducting platforms by extending the preparation of two-qubit Bell states to multipartite entanglement [34]. Further demonstrating their viability in real-world quantum systems, subsequent experiments have demonstrated reliable multi-qubit GHZ-state creation and detection in circuit QED designs [35].

Protocols for the resilient distribution of GHZ states have been presented, along with approaches for generating them. Interestingly, loss-tolerant techniques enable effective distribution of entanglement over quantum networks, thereby improving their suitability for practical communication scenarios [36]. Using high-fidelity digital quantum circuits, more recent developments have shown scalable techniques for creating, sustaining, and modifying large-scale GHZ entanglement. Significant progress towards large-scale quantum systems has been achieved thanks to these methods, which have enabled the creation of genuinely entangled GHZ states involving up to 60 qubits during initialization [37].

The vulnerability of multipartite entanglement to decoherence is a significant feature. For systems up to $N = 15$ qubits, experimental investigations show a linear dependence of the decoherence rate on the number of qubits, with $\alpha = (7.13N + 5.54) \times 10^{-3}\ \mu s^{-1}$, confirming the lack of super-decoherence effects. Additionally, coherence has significantly improved with the use of dynamical decoupling techniques, especially for intermediate-scale systems like 7-qubit GHZ states. Large-scale entanglement has also been achieved on superconducting quantum processors in the form of completely bipartite entangled native graph states, in addition to GHZ states. Specifically, experiments on IBM quantum devices, such as the IBM Osprey 433-qubit processor, have demonstrated entanglement over up to 414 qubits, representing a significant advancement in scalable quantum hardware [38].

## C. W state circuit

The set of truly trifold entangled states is divided into two sets that are unrelated under local operations assisted with classical communication (LOCC) when any nontrivial tripartite entangled state is transformed using stochastic local operations and classical communication (SLOCC) into one of two standard forms, either the GHZ state or another state as the W state [39]. For an n-qubit system, the generalized form is (3) as given by:

$$\frac{1}{\sqrt{n}} \sum_{i=1}^{n} |e_i\rangle \qquad (3)$$

Where the i-th qubit is in state $|1\rangle$ and every other qubit is in state $|0\rangle$, the symbol $|e_i\rangle|$ indicates a computational basis state. Fault-tolerant quantum computers repeatedly apply a four-step process. Iterative error-correction cycles are essential for scalable quantum computation. These usually entail four steps: (i) a series of one- and two-qubit quantum gates are executed; (ii) syndrome measurement is performed on specific qubits; (iii) classical processing is used to detect the existence and location of errors; and (iv) corrective operations are applied to reduce these errors. The foundation of dependable large-scale quantum computation is formed by repeating this cycle with successive gate operations [40].

Extensions to higher-dimensional quantum systems (qudits) have also been investigated in addition to qubit systems. Specifically, for systems with prime power dimensions, controlled-Hadamard (controlled H) gate designs in odd qudit dimensions allow the creation of arbitrarily large W states, generalizing multipartite entanglement to higher-dimensional Hilbert spaces [41].

The creation and characterization of high-order W states have advanced significantly experimentally. Precise state characterization methods and the creation of such states with great accuracy have been shown by integrated quantum circuit platforms [42]. These developments demonstrate that complicated entangled states can be implemented in scalable quantum technology.

More generally, massive quantum systems exhibit multipartite entanglement in various inequivalent forms, requiring the creation of states from various entanglement classes based on the intended use. W states are especially important in quantum networking protocols because of their robustness and entanglement distribution properties, which make them ideal for tasks such as resource sharing across quantum networks and multi-party communication [43].

## D. Many qubit systems of circuits

The capacity to use quantum algorithms on multi-qubit devices has been crucial to the development of large-scale quantum computation. A three-qubit nuclear magnetic resonance (NMR) quantum computer was used in early experimental demonstrations to ascertain the periodicity of an input state via the quantum Fourier transform. These implementations are fundamental milestones in the realization of algorithms such as Shor's algorithm, in which QFT is a key computational primitive [44].

Circuit initialization and entanglement methods become more crucial as quantum technology grows. For example, in a 127-qubit system, quantum circuits can be initialized by dividing qubits into clusters, each of which consists of a subset (for example, ten qubits) where QFT is applied independently, and entanglement is established. This clustering technique maintains the capacity to perform frequency-domain manipulations on quantum states while allowing for controlled scaling [45].

Studies have investigated the application of Shor's factorization algorithm on superconducting platforms, namely Josephson-junction-based charge qubit registers, thereby bridging the gap between algorithmic design and physical realization. Factoring the number 21 is one example of how algorithmic decomposition into two- and three-qubit gates can be optimized for hardware with short coherence times. Gate implementations are customized using numerical optimization to ensure compatibility with physical constraints such as connections and decoherence [46].

Quantum circuit compilation and layout optimization are crucial at the software-hardware interface. Comprehensive circuit layout information, including permutation mappings introduced during transpilation, is provided by frameworks like Qiskit. Initial layouts, which allocate logical qubits to physical hardware, and final layouts, which account for further transformations, such as SWAP gate insertions required for routing on restricted architectures, comprise these mappings [47].

The Quantum Approximate Optimization Algorithm has been extensively studied as a near-term quantum algorithm for solving combinatorial optimization problems, such as Max-Cut, in which each graph node is mapped to a qubit in a multi-qubit system [48]. These algorithms' efficacy is directly related to the coherence and connectivity characteristics of the underlying hardware.

To enable scalability, contemporary superconducting quantum processors are arranged into hierarchical designs. For instance, arrays of 8 qubits per 4x4 unit cell, coupled by tunable couplers, have been created. These architectures take into account practical factors, such as dormant qubits undergoing calibration, and provide specific connectivity patterns between qubits, with edges denoting configurable coupling strengths [49].

Combinatorial optimization problems involving up to 127 qubits have been successfully solved by comprehensive quantum solvers built on such systems. Interestingly, Max-Cut instances on random regular graphs have been successfully solved for systems close to 120 qubits, even though the hardware connectivity limitations and the problem topology are not exactly aligned [50].

The system size has steadily increased as superconducting quantum technology has advanced; qubit counts have increased from 127 to 433 and beyond, with predictions approaching thousands of qubits. Key paradigm shifts in quantum computing, such as the transition from Noisy Intermediate-Scale Quantum (NISQ) devices to early fault-tolerant quantum computing (EFTQC) and finally to completely fault-tolerant quantum computing (FTQC), are supported by this trend. These transitions are distinguished by their ability to encode multiple logical qubits using physical qubit resources and to perform fault-tolerant logical operations, such as non-Clifford gates [51].

Lastly, improvements in control have further increased the effectiveness of quantum computing. By directly exposing calibrated pulses as logical gates, methods such as cross-resonance pulse engineering enable pulse-efficient circuit implementations on noisy hardware. Without needing in-depth user knowledge of low-level pulse control, this method enhances gate quality, minimizes the requirement for additional transpilation overhead, and enables better use of limited coherence times [52].

## E. Error correction, mitigation, and work contribution

One of the most important findings in practical quantum hardware is that computing reliability is directly impacted by circuit complexity. Specifically, the number of logical gates in a bit sequence determines how many quantum gates are applied to it, thereby affecting execution fidelity and overall error accumulation [53]. This emphasizes how circuit form and depth affect near-term quantum devices.

To overcome these difficulties, methods such as quantum error mitigation have become viable ways to improve computational results in noisy quantum systems. Error mitigation, as opposed to full error correction, focuses on using post-processing techniques to reduce the impact of noise. Prominent techniques that mimic device noise and modify measurement results are Zero Noise Extrapolation (ZNE) and Probabilistic Noise Cancellation (PNC). These methods are especially useful for Noisy Intermediate-Scale Quantum (NISQ) systems because they improve final output accuracy without requiring extra qubits. By embedding quantum information into logical qubits and actively correcting computation faults, quantum error correction seeks to preserve quantum information. Fault-tolerant qubits backed by reliable error-correcting protocols will eventually be needed to achieve scalable quantum advantage, requiring improvements in both software and hardware solutions [54].

Such methods have been developed and tested in large part because to early software frameworks. For instance, Qiskit Ignis enabled the modular construction of new error-correcting codes and decoding techniques by introducing implementations of repetition codes within its topological_codes module [55]. Experimentation with fault-tolerant schemes in simulated and real-device contexts has been made easier by these technologies.

The nature of quantum information itself presents a fundamental problem for quantum error correction. Qubits are more vulnerable to a wider range of errors, such as phase flips and decoherence, than classical bits, which mostly encounter bit-flip faults. As a result, many error channels must be considered by quantum error-correcting codes. Additionally, the system usually collapses when quantum states are directly measured, requiring the use of supplementary qubits and non-demolition syndrome measurements to identify and fix faults without affecting the encoded data. The complexity of creating fault-tolerant quantum systems is greatly increased by this requirement [56].

More broadly, the variety of quantum computing technologies introduces new factors to consider. Comparative analysis is crucial because different hardware platforms exhibit distinct noise characteristics, connectivity limitations, and operational fidelity. Confidence in achieving a useful quantum advantage is rapidly increasing due to ongoing improvements in precision and processing speed, bolstered by cooperative efforts across industry and research consortia. Effective quantum computing, however, is still a long way off, requiring years of consistent development and domain knowledge [54].

In this regard, widely used quantum circuits—such as the W state, the Greenberger–Horne–Zeilinger state, and the Quantum Fourier Transform—serve as fundamental building blocks for more intricate quantum applications. These circuits can be scaled up to larger qubit systems when implemented separately or in tandem, making them more vulnerable to noise in NISQ settings. To increase trust in the outcomes of actual quantum technology, error mitigation and correction mechanisms must be incorporated.

Using both Qiskit simulators and IBM's 127-qubit Sherbrooke processor, this work advances the methodical modeling and analysis of QFT, GHZ, and W-state quantum circuits across a range of 4 to 10 qubits. To evaluate hardware compatibility, the study looks at differences in circuit layout factors as depth, width, and qubit count. To facilitate direct comparison, each circuit design is run with 4096 shots on both the simulator and the quantum processing unit (QPU) samplers.

This work's novelty lies in comparing 21 baseline quantum circuits with their transpiled counterparts, both in simulation and on real superconducting hardware, in the NISQ regime. The study contributes to the potential to incorporate error-correction and mitigation techniques into realistic quantum workflows and offers insights into the effects of hardware limitations by using quantitative metrics to examine performance differences across various qubit configurations.

# III. Approach- Circuits in the Simulator and QPU

Equations (1), (2), and (3) were used to build three classes of n-qubit circuits for qubit sizes ranging from 4 to 10 within the Qiskit ecosystem: the Quantum Fourier Transform, Greenberger–Horne–Zeilinger state, and W state. The statevector simulation framework was used to get initial state representations, and each circuit was run for 4096 shots to provide corresponding state distributions. Quantum states before and after measurement were qualitatively compared using the Bloch sphere representation.

Each circuit was then run for 4096 shots on the Aer simulator before being hardware-awarely transformed using the transpilation method to guarantee compatibility with superconducting quantum hardware. The IBM Sherbrooke 127-qubit processor, chosen for its availability and comparatively short queue time among available quantum processing units, was used to analyze transpiled circuits using the Sampler primitive.

Several statistical distance measures, such as Total Variation Distance (TVD), Kullback–Leibler (KL) divergence, Jensen–Shannon divergence, and Hellinger distance, were used to quantitatively evaluate the agreement between simulated and hardware-executed results. By evaluating distributional similarity, these measures show how well circuit outputs retain significant quantum information in various execution contexts.

These metrics were used to determine the ideal qubit arrangement for each circuit class, and bar charts were used to facilitate comparison analysis. To evaluate consistency across execution modes, the analysis included comparisons with the Estimator primitive and Sampler outputs using resilience approaches. Furthermore, certain optimum circuits were subjected to an error-mitigation method based on dynamical decoupling, and the ensuing performance gains were examined.

After 21 circuit cases were run on both simulators and actual superconducting hardware, error mitigation strategies were specifically used. With a focus on trade-offs between circuit design complexity and hardware performance, the findings were analyzed to compare baseline and transpiled circuit properties. In addition to highlighting potential future research directions and the limitations of the current study, these findings offer insights into circuit optimization methodologies under practical hardware constraints.

## IV. Results and Discussion

Using Qiskit, 4-qubit quantum circuits for QFT, Fig. 1, 2, and 3, the GHZ state, and the W state, depict the use of respective applicable gates and measurements. The corresponding circuits were transpiled with the IBM_Sherbrooke backend, as described in Figs. 4, 5, and 6, respectively, showing the difference in resulting layouts for mapping onto real quantum hardware. The circuit depth and width are adjusted during transpilation to match the physical qubits.

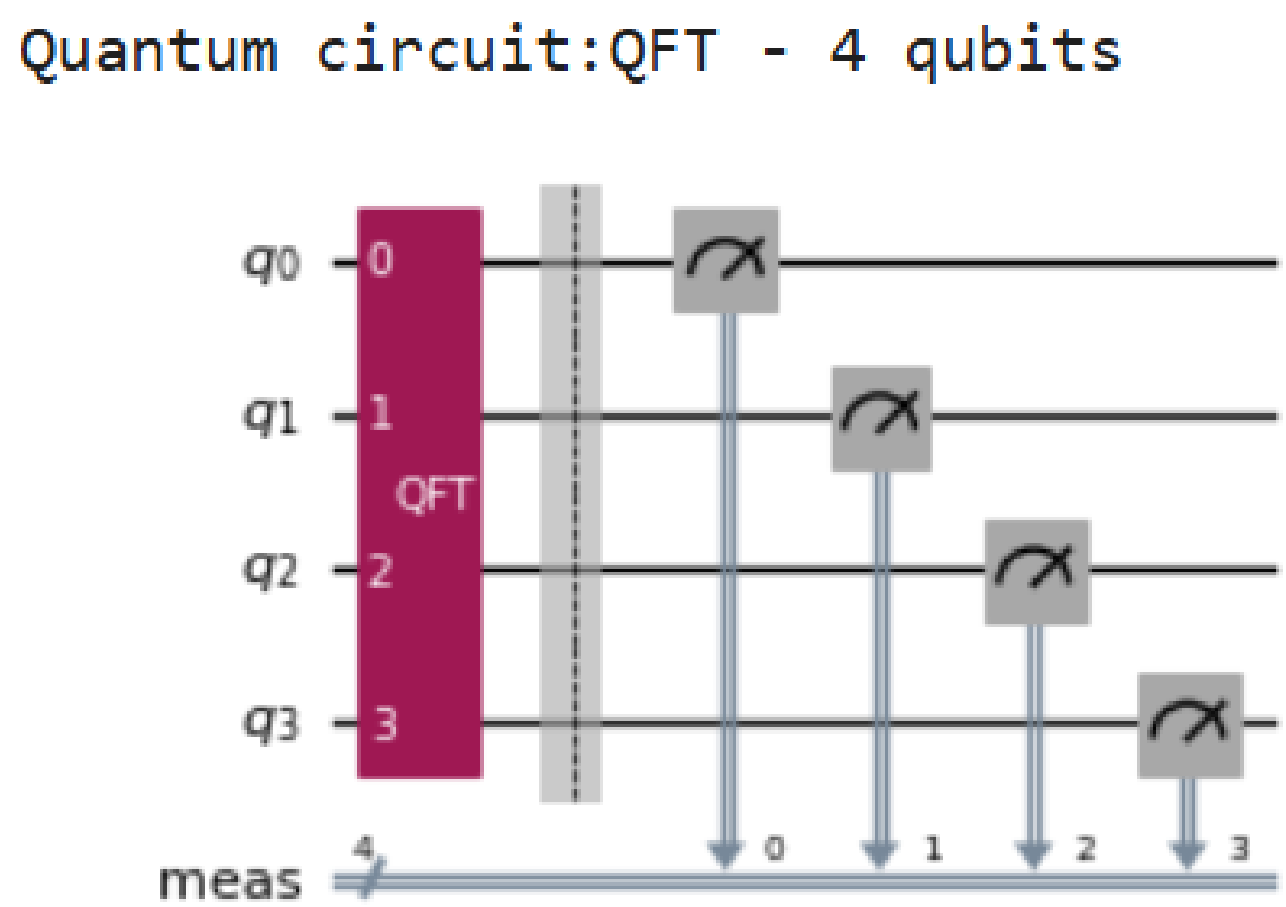

*Fig. 1: Quantum circuits with 4 qubit: QFT*

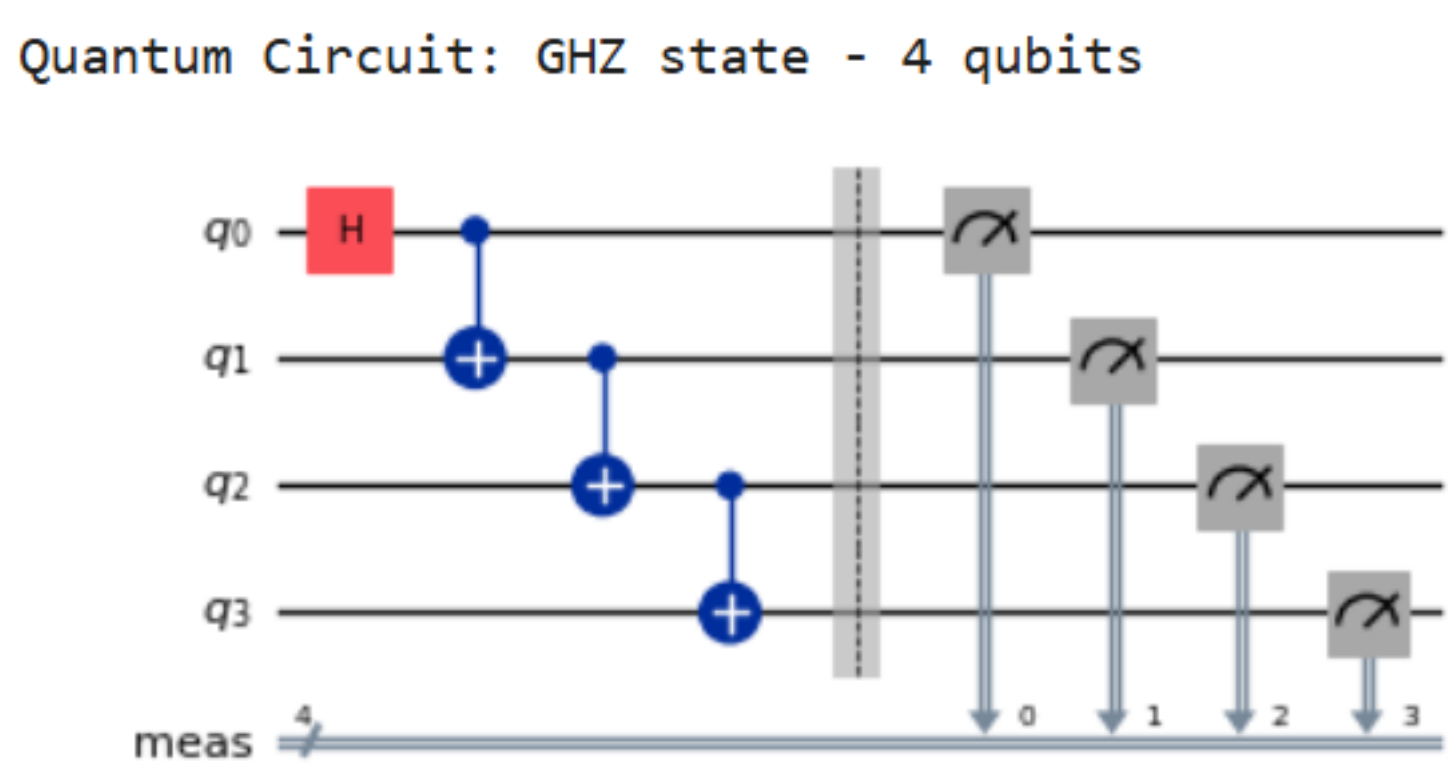

*Fig. 2: Quantum circuits with 4 qubit: GHZ*

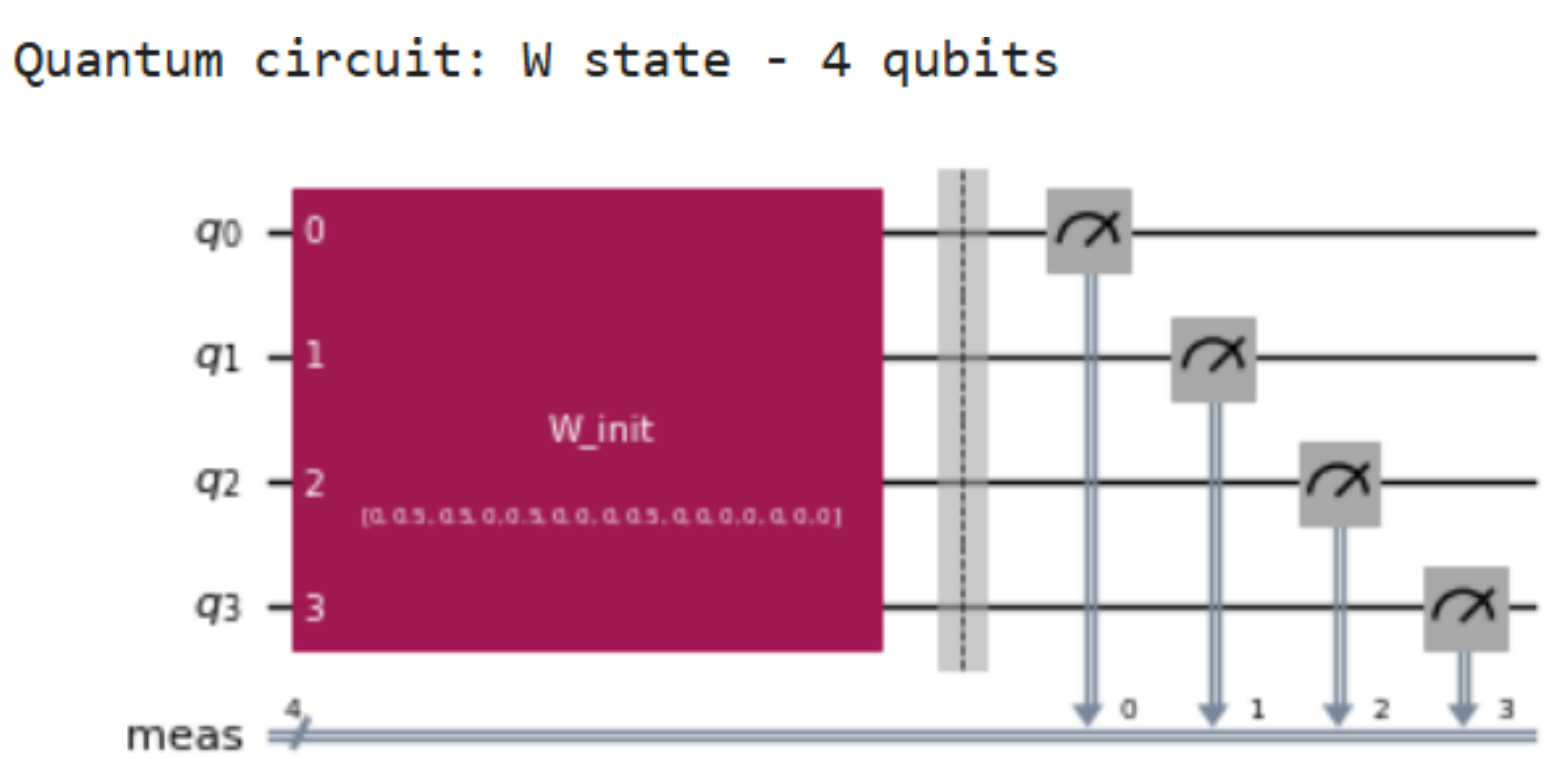

*Fig. 3: Quantum circuits with 4 qubit: W state*

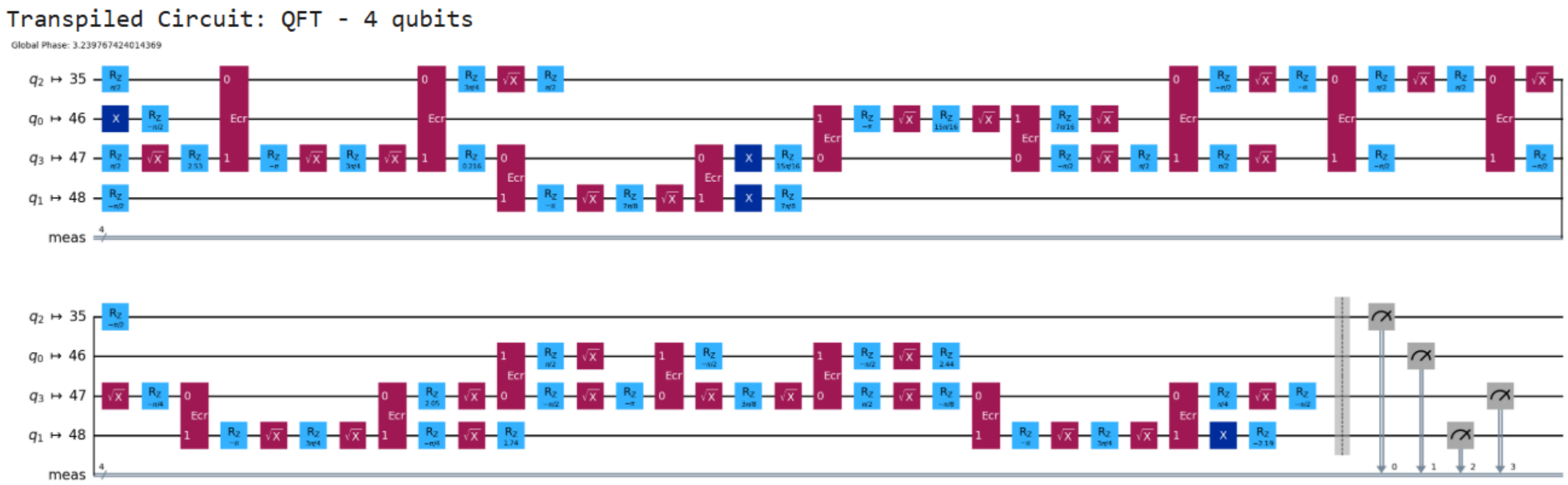

*Fig. 4: Transpiled Quantum circuits with 4 qubit: QFT.*

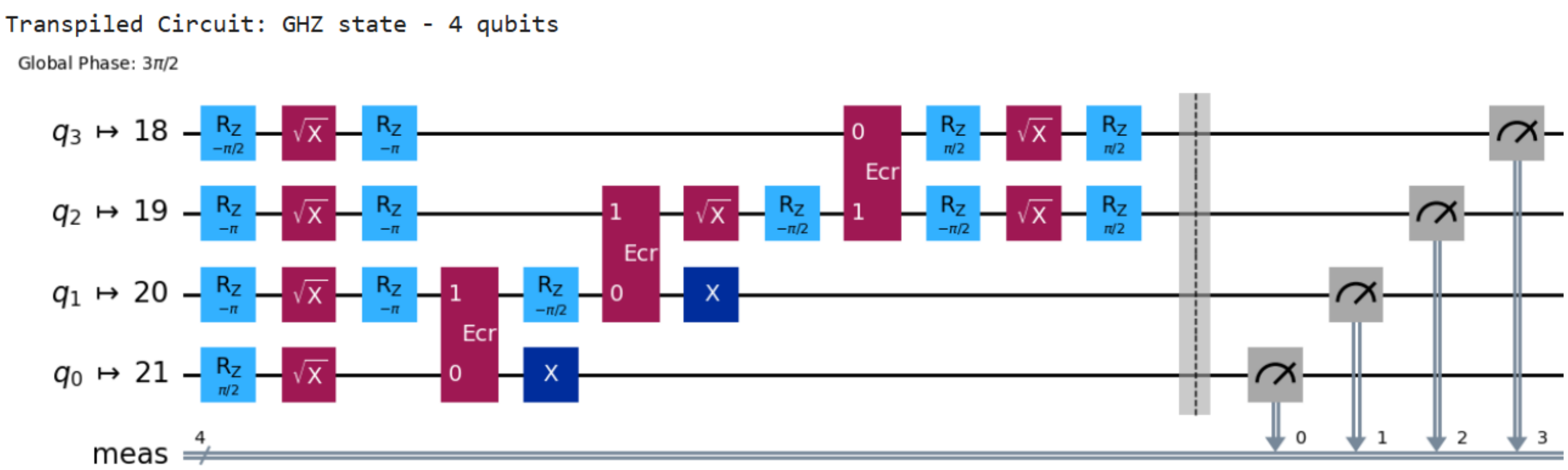

*Fig. 5: Transpiled Quantum circuits with 4 qubit: GHZ*

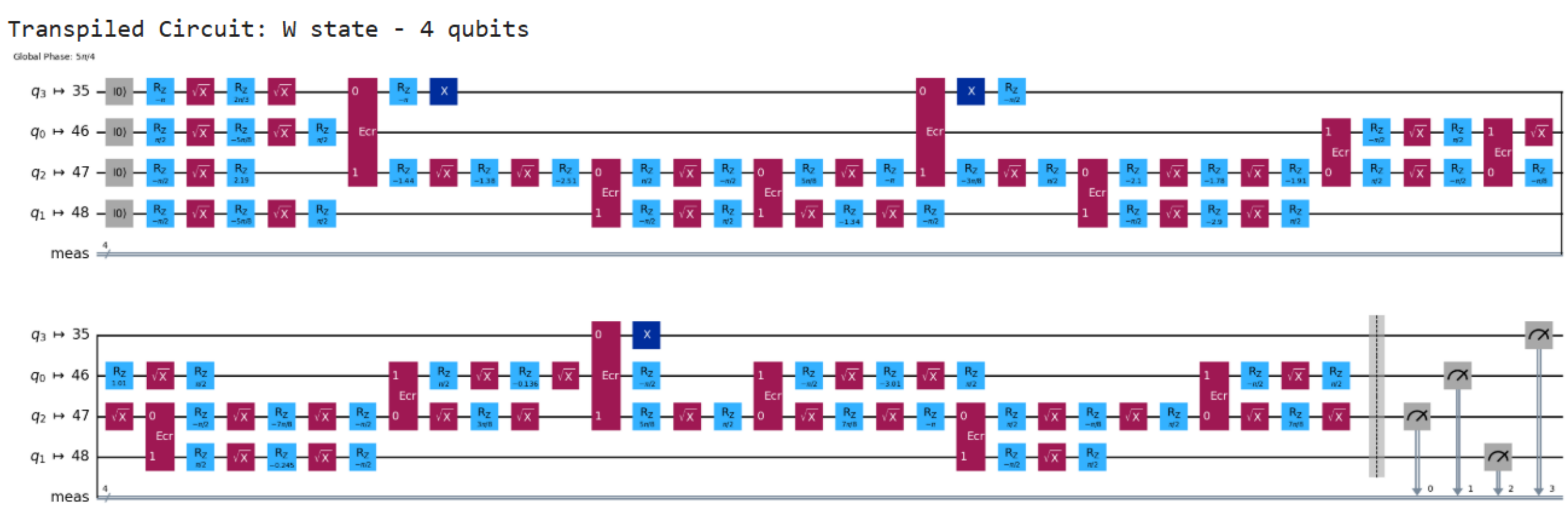

*Fig. 6: Transpiled Quantum circuits with 4 qubit: W state.*

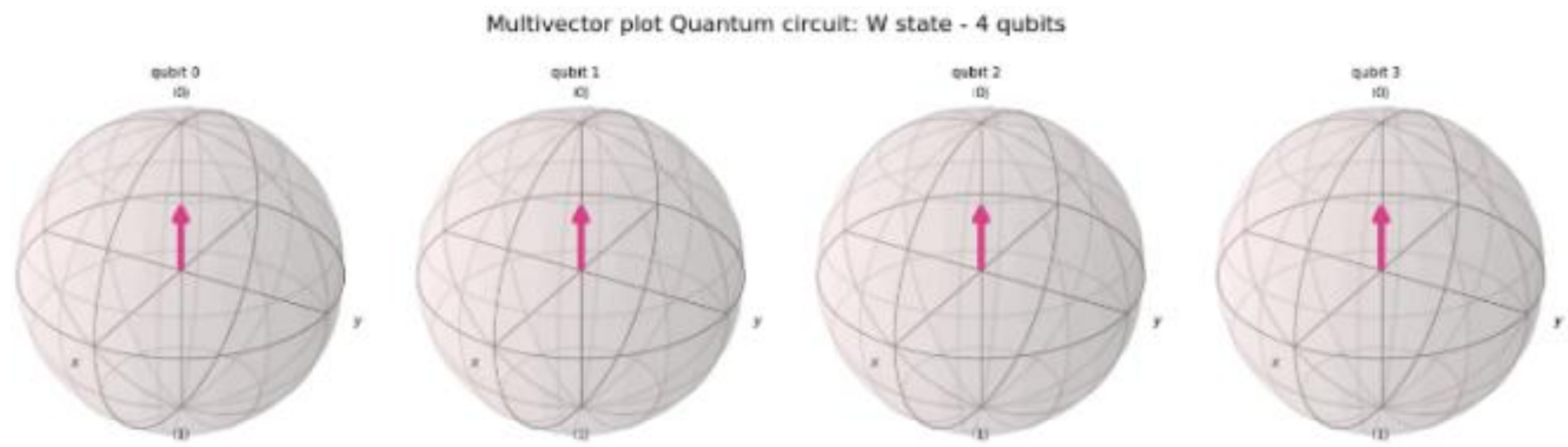

*Fig. 7: 4-qubit W state multi-vector plot from statevector before measurement*

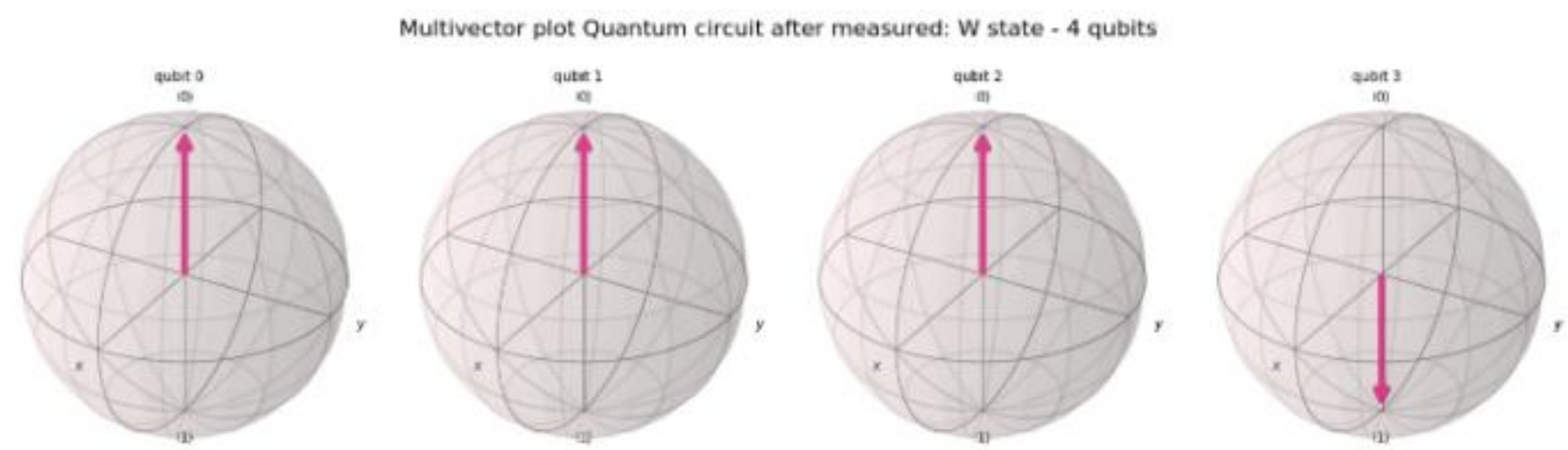

*Fig. 8: 4-qubit W state multi-vector plot from state vector after measurement.*

Fig. 7 and 8 depict the 4-qubit W state before and after measurement, respectively, using the Aer Statevector simulator after 4096 shots. The red arrow, initial state, Fig. 7, indicates the state vectors pointing towards the Z direction with no superposition or entanglement; however, shorter lengths resemble a mixed state in the |0⟩ state, Fig. 8. In comparison, the post-measured classical state of qubit 3 is evident from the red arrow pointing towards state |1⟩, indicating a difference in coherence from the pre-measure state.

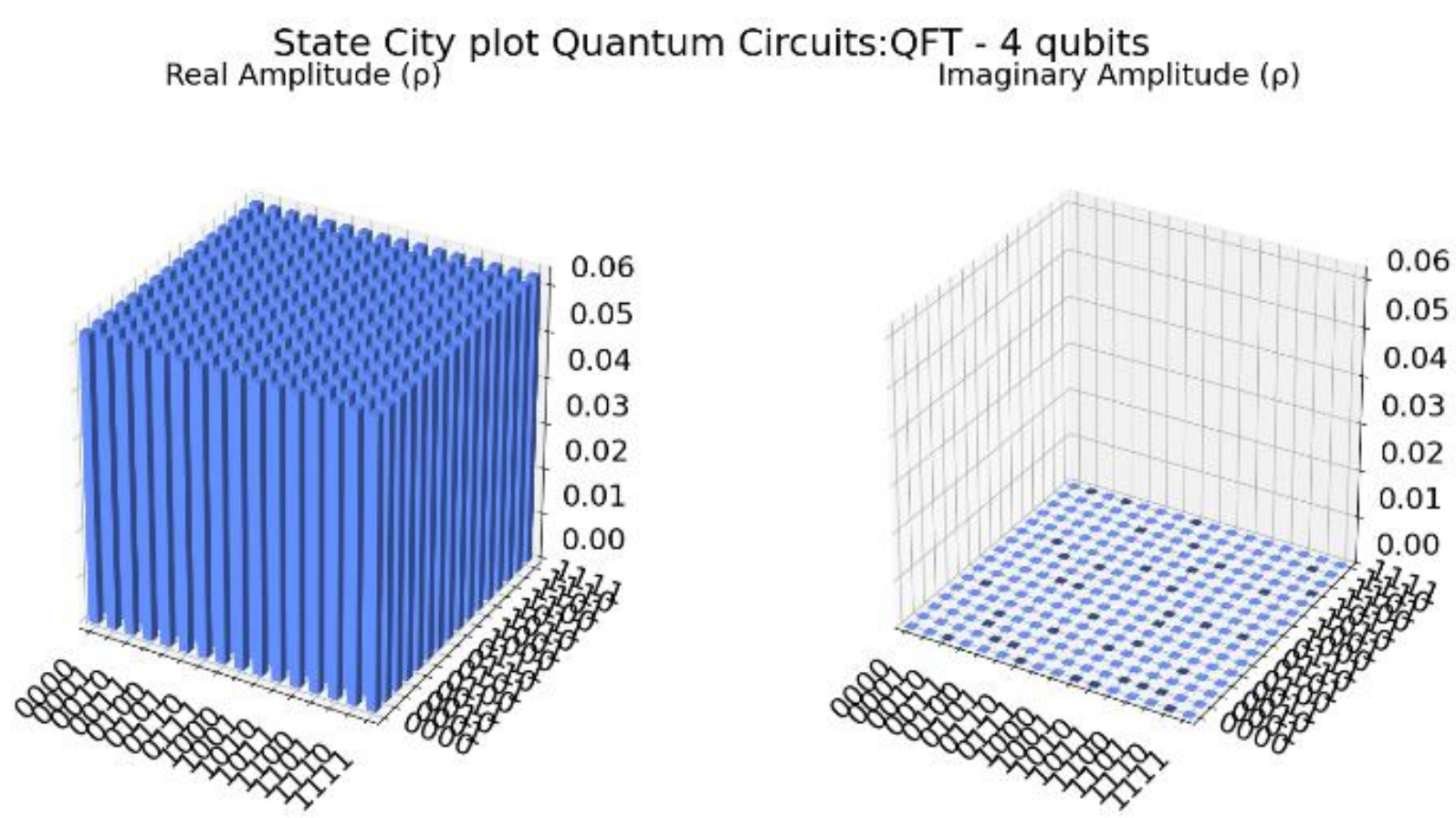

*Fig. 9: 4-qubit QFT state city plot from state vector before measurement.*

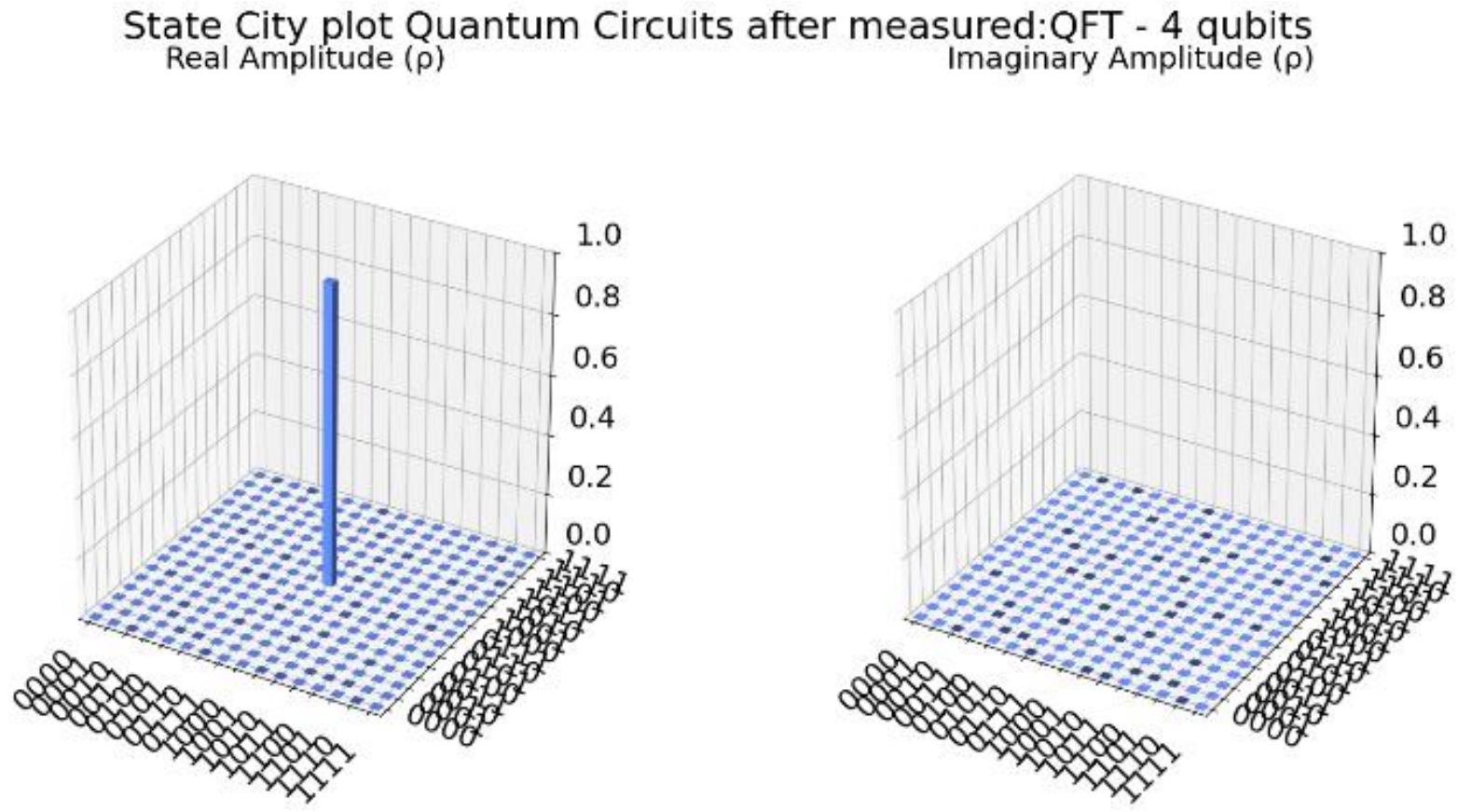

*Fig. 10: 4-qubit QFT state city plot from the state vector after measurement.*

Fig. 9 shows a 4-qubit QFT system, where each state has a probability of 0.06 before measurement, which collapses to the classical state |0001⟩ after measurement (Fig. 10).

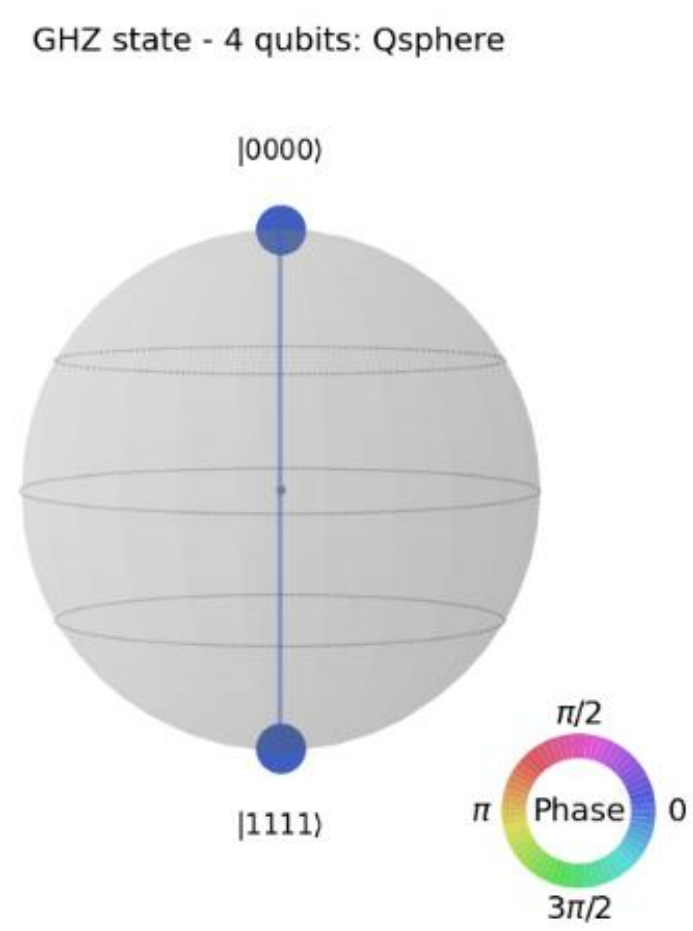

*Fig. 11: GHZ state – 4 qubit Qsphere from statevector before measurement.*

Fig. 11 and 12 depict a 4-qubit GHZ state, a classic pure superposition, and a highly entangled state that collapses to |1111⟩, respectively.

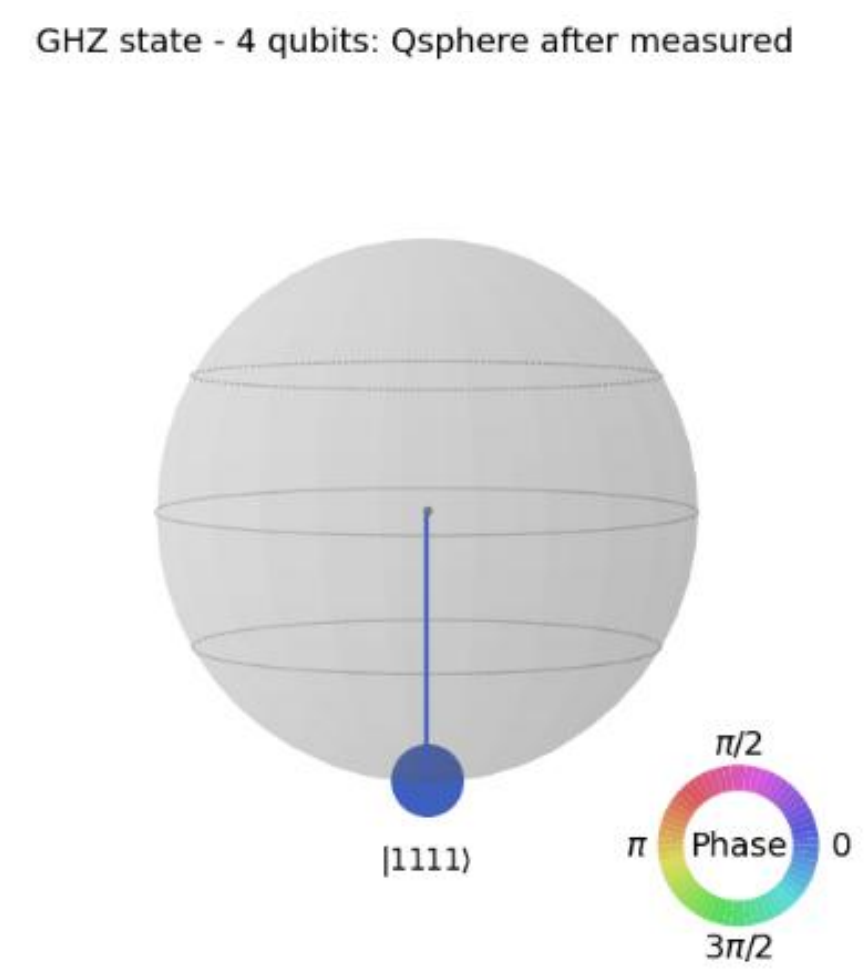

*Fig. 12: GHZ state – 4 qubit Qsphere from statevector after measurement*

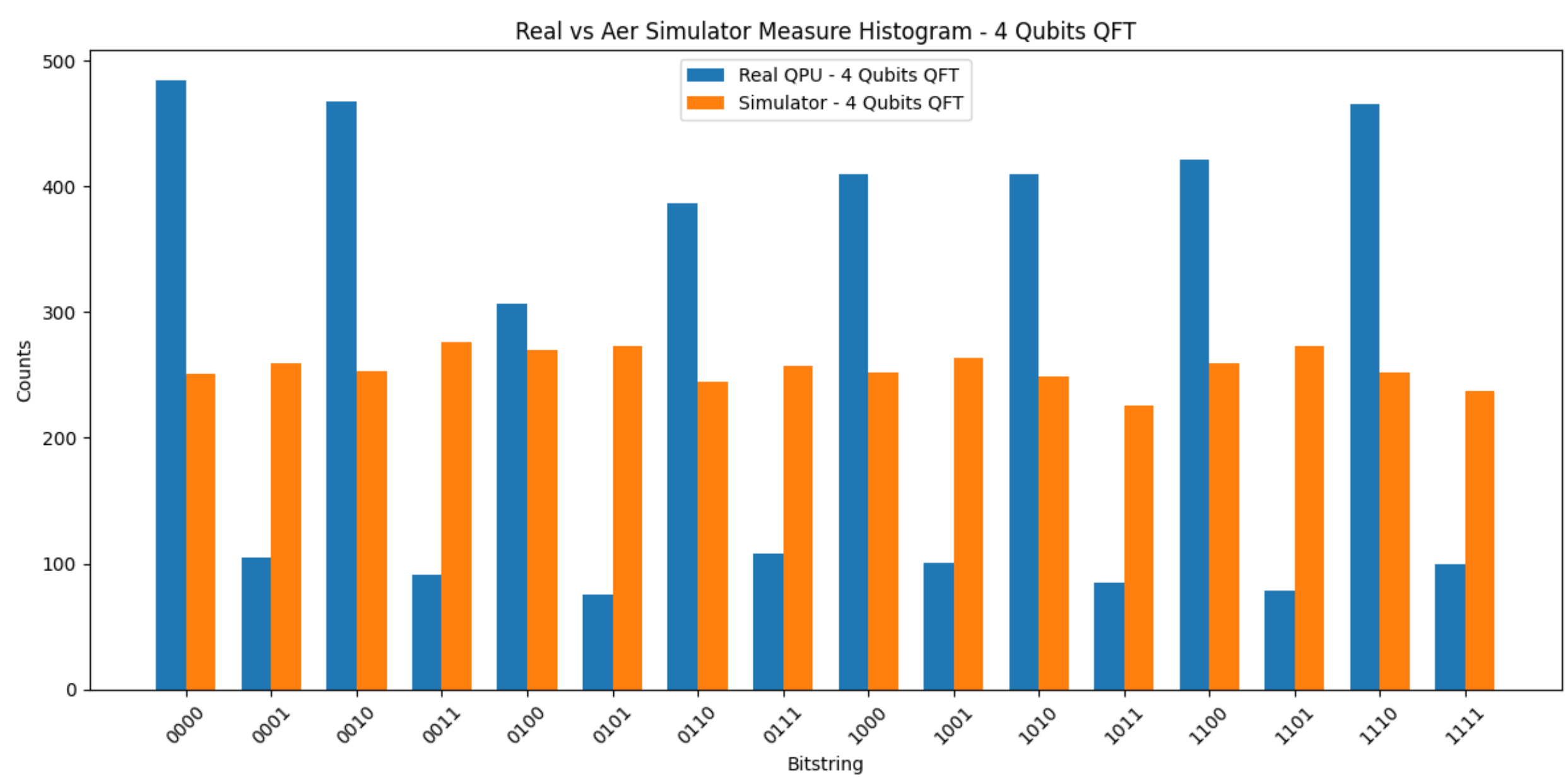

.

*Fig. 13: 4 Qubit QFT Aer Simulator and IBM Sherbrooke histogram.*

Figure 13 compares Aer Simulator and IBM Sherbrooke count results for 4096 shots of each 4-qubit QFT circuit, post-measurement. The simulator results are mostly uniform, while the taller and shorter bars for some states in the QPU count indicate differences.

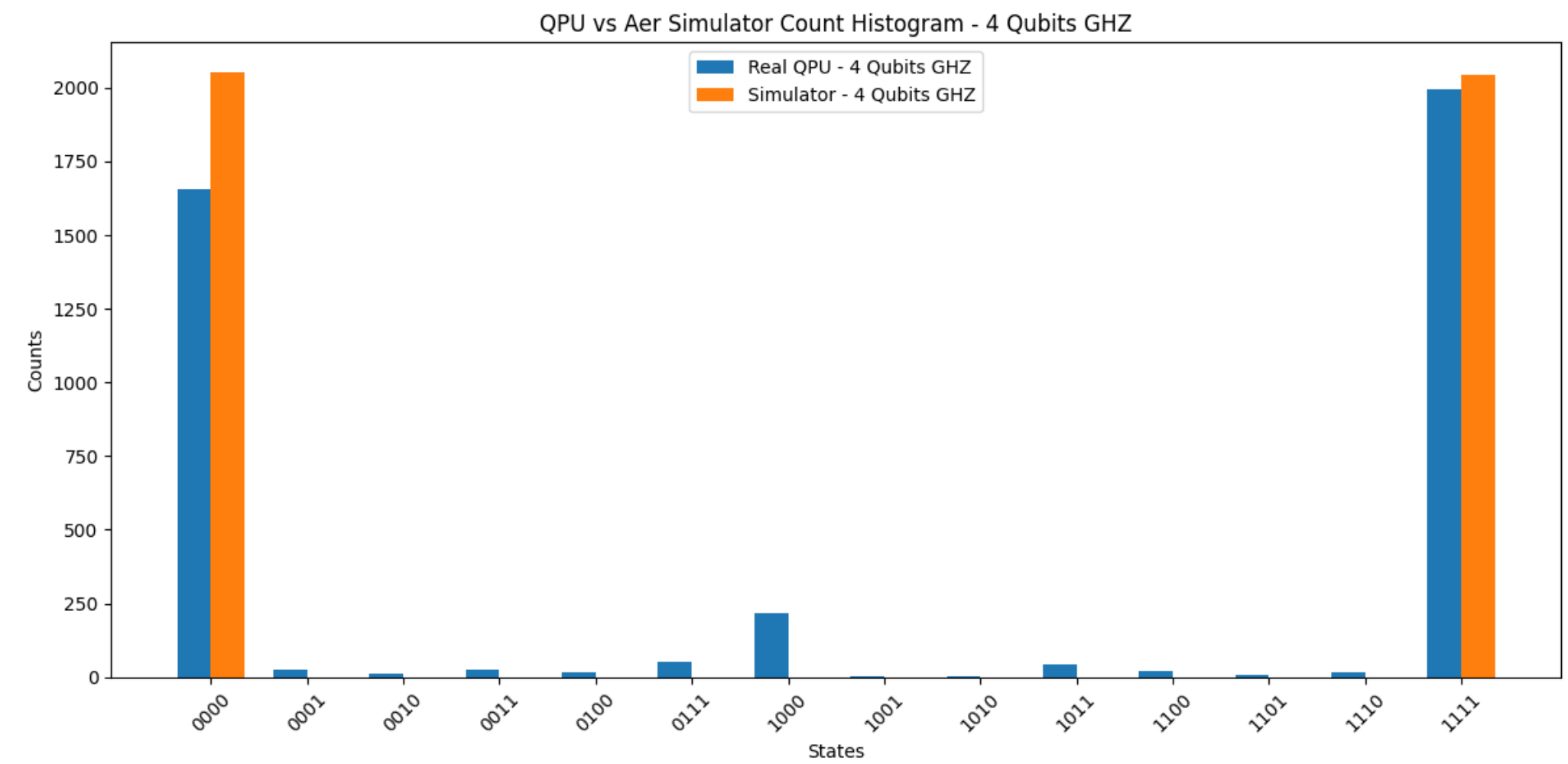

*Fig. 14: 4 Qubit GHZ Aer Simulator and IBM Sherbrooke histogram*

Fig. 14 reveals the stark difference in simulator and QPU counts with the introduction of noise in shorter blue bars, while the simulator results are consistent for 4-qubit GHZ states.

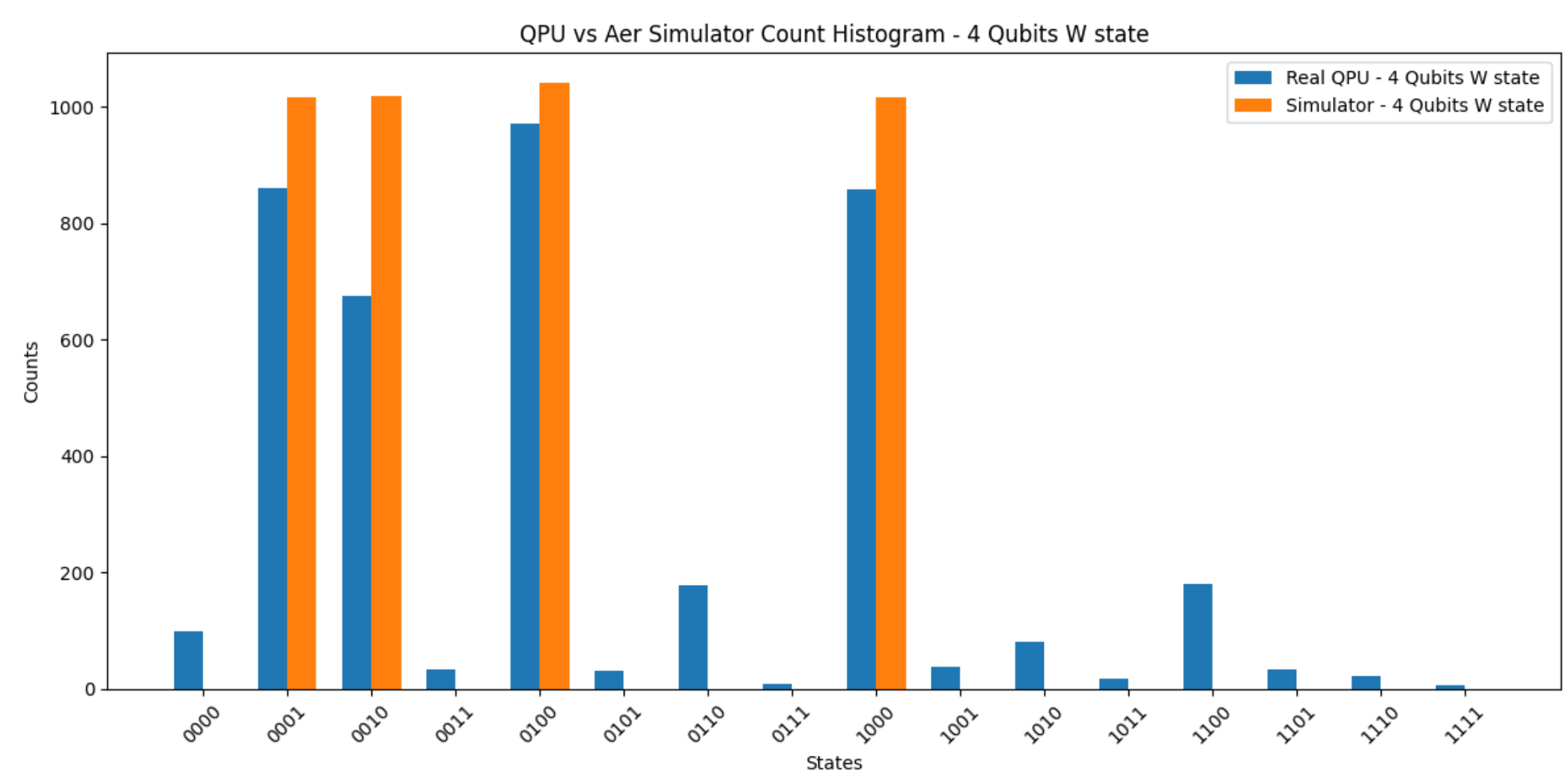

*Fig. 15: 4-qubit W-state Aer Simulator and IBM Sherbrooke histogram*

Fig. 15 depicts a 4-qubit W state in a symmetric superposition, with each |1⟩ state having nearly equal counts as 1 qubit, while the others are in |0⟩, as per Aer Simulator results. This differs from the QPU results due to noise in other states, as indicated by the short blue bars. Therefore, as shown in the above Fig., QPU runs for 4-qubit systems yield noise in the QPU outcomes compared with the simulator results. In this follow-up, the circuits are run through subsequent qubit systems, viz., 5 to 10, and the comparative results further indicate significance.

*Table 1: Qubits to Classical and Quantum Fourier Transform Comparison*

| **Qubits** | **Classical FFT Vectors** | **Quantum QFT Qubits** |
|---|---|---|
| 3 | 8 complex numbers | 3 |
| 10 | 1024 complex numbers | 10 |
| 20 | 1 million+ entries | 20 qubits |
| 30 | 1+ billion amplitudes | 30 qubits |

Table 1 lists 2^n basis states, where n is the number of qubits, corresponding to amplitudes that can be represented simultaneously in the quantum Fourier transform due to superposition. This holds for 30 qubits, up to a billion, a stark contrast with the limitations of classical computation, highlighting the exponential growth. Table 2 further illustrates that the RAM requirement for qubits increases: 30 qubits require 16 GB and 37 qubits necessitate an astonishing 2048 GB, underscoring the need for quantum processors.

*Table 2: Required RAM size against qubits*

| | **RAM Limit (GB)** | **Qubits** |
|---|---|---|
| 0 | 16 | 30 |
| 1 | 32 | 31 |
| 2 | 64 | 32 |
| 3 | 128 | 33 |
| 4 | 256 | 34 |
| 5 | 512 | 35 |
| 6 | 1024 | 36 |
| 7 | 2048 | 37 |

Table 2 further illustrates that the RAM size requirement for qubits increases, which for 30 qubits requires 16 GB, and for 37 qubits, astonishingly, 2048 GB, necessitating a simulator size that matches, rather than justifies, the use of quantum processors.

The quantum circuit outcomes from 4 to 10 qubits have been plotted on a logarithmic scale for comparison. Fig. 16 shows that the metrics saturate or stabilize beyond six qubits, except for the KL distance. Individually, for QFT - six qubits, GHZ – 5 qubits, and W state – 4 qubits exhibit all four metrics at their lowest compared to others. This implies that the simulation and QPU results closely resemble the outcome, with the threshold for the individual quantum circuits requiring almost no error correction. However, for QFT and GHZ above 9 qubits, the graphs tend to converge except for the W state, where the stabilization reaches from 6 qubits. The scope of error correction using available techniques in the IBM cloud, namely dynamic decoupling, gate twirling, and M3 techniques, appears to be the most effective for improving the circuit's outcome.

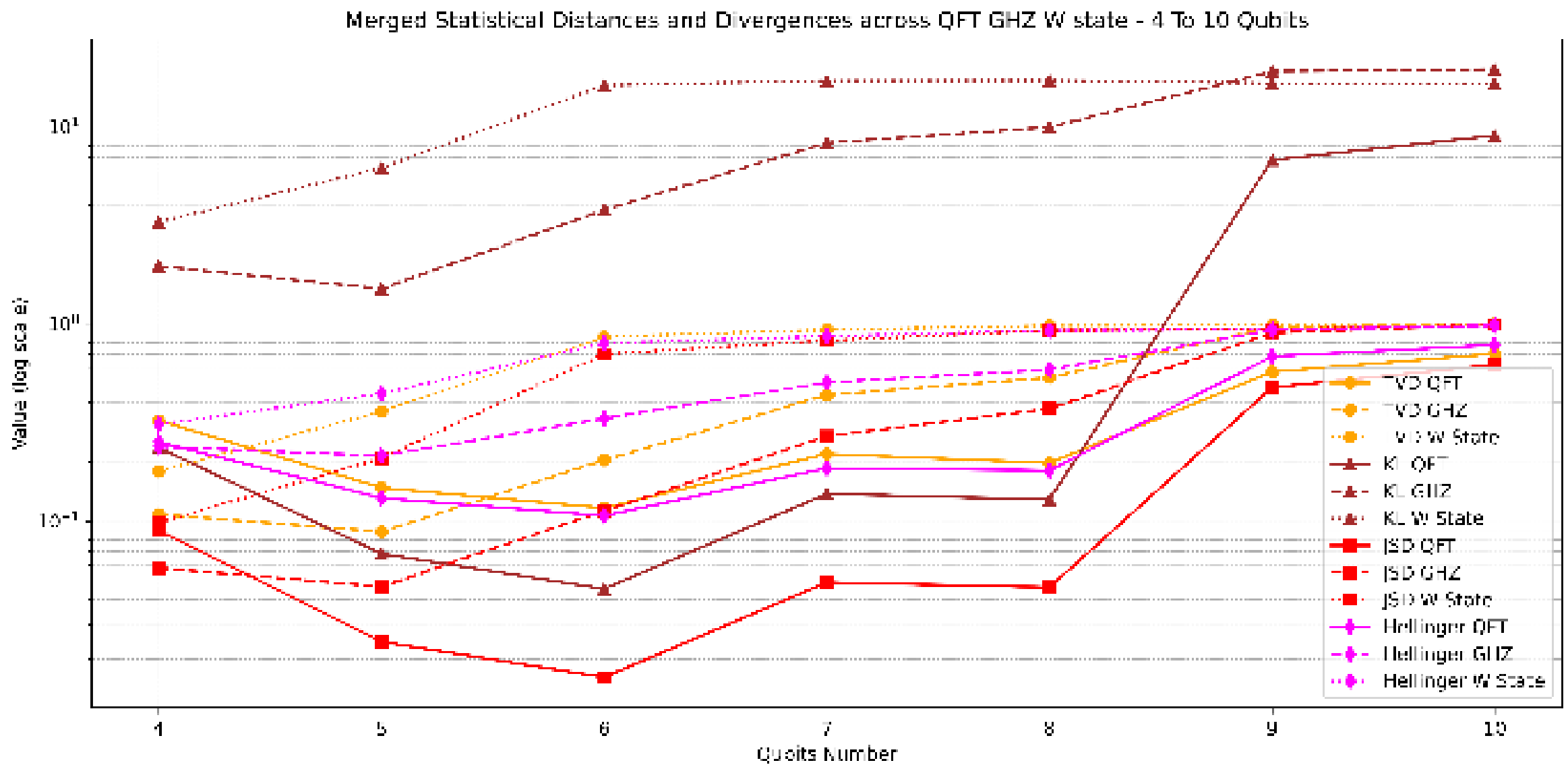

*Fig. 16: QFT, GHZ & W state – 4 to 10 qubits - Different metrices plot on qubit number and values in log scale*

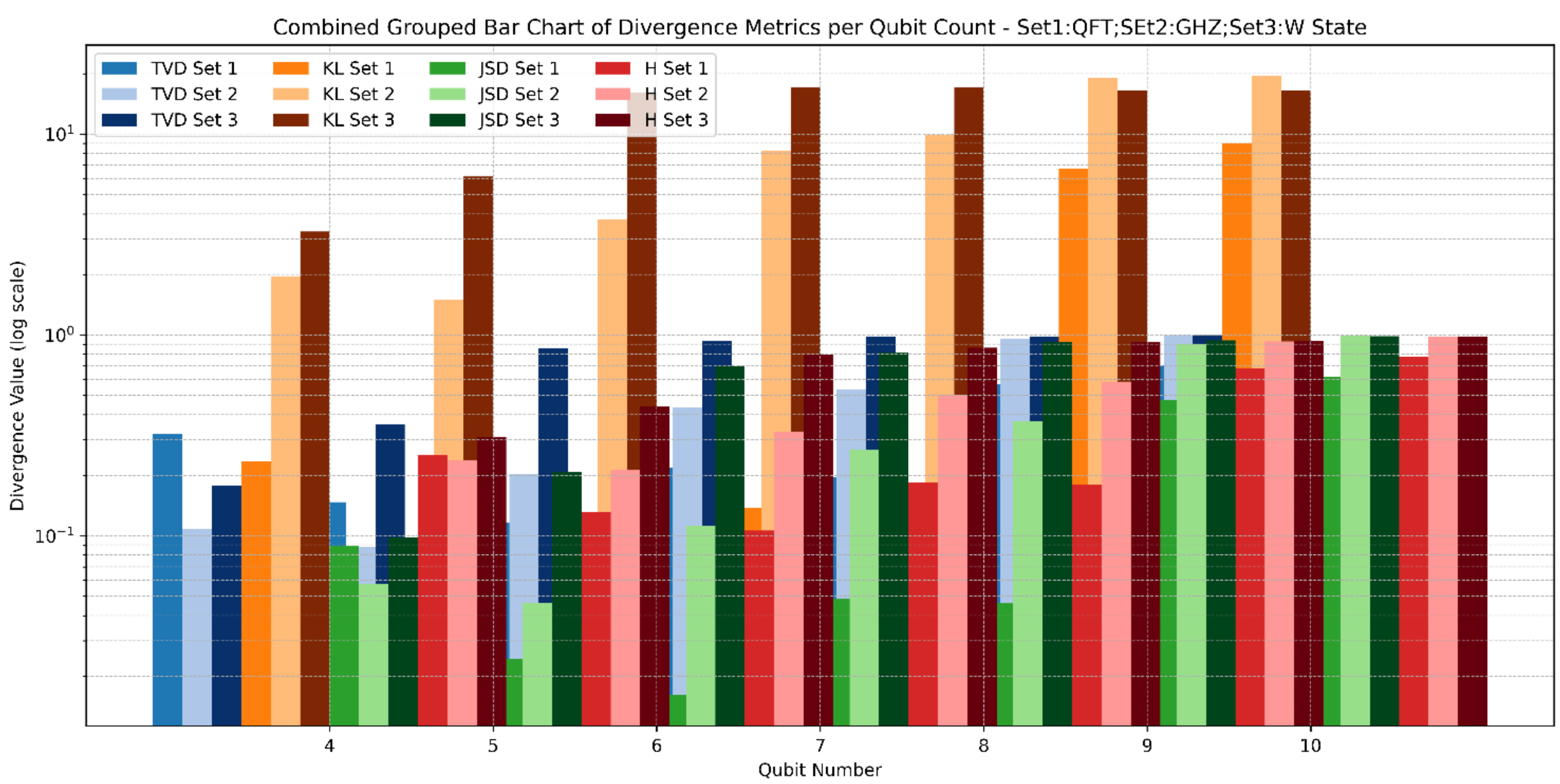

*Fig. 17: Different Metrics comparison plot.*

Based on Fig. 16, the metric values calculated for QFT–6 qubits, GHZ–5 qubits, and W state 4 qubits are produced in the subsequent Fig. 17. The Hellinger distance for QFT–6 qubits appears the minimum of 0.016, implying proximity of results in the aer simulator and QPU. These levels are considered identical for comparison: the Sampler without error correction compared with the Estimator, a better candidate for adoption in a QPU run.

*Table 3: QFT quantum and transpiled circuit details*

| circuit | depth | width | measurem ents | gates | barriers | no. of qubits |
|---|---|---|---|---|---|---|
| **Quantum circuit** | | | | | | |
| 4 | 2 | 8 | 4 | 1 | 1 | 4 |
| 5 | 2 | 10 | 5 | 1 | 1 | 5 |
| 6 | 2 | 12 | 6 | 1 | 1 | 6 |
| 7 | 2 | 14 | 7 | 1 | 1 | 7 |
| 8 | 2 | 16 | 8 | 1 | 1 | 8 |
| 9 | 2 | 18 | 9 | 1 | 1 | 9 |
| 10 | 2 | 20 | 10 | 1 | 1 | 10 |
| **Transpiled quantum circuit** | | | | | | |
| 4 | 75 | 131 | 4 | 103 | 1 | 127 |
| 5 | 102 | 132 | 5 | 168 | 1 | 127 |
| 6 | 150 | 133 | 6 | 283 | 1 | 127 |
| 7 | 234 | 134 | 7 | 406 | 1 | 127 |
| 8 | 251 | 135 | 8 | 529 | 1 | 127 |
| 9 | 285 | 136 | 9 | 675 | 1 | 127 |
| 10 | 401 | 137 | 10 | 980 | 1 | 127 |

*Table 4: GHZ quantum and transpiled circuit details*

| circuit | depth | width | measure ments | gates | barriers | no.of qubits |
|---|---|---|---|---|---|---|
| Quantum circuit | | | | | | |
| 4 | 5 | 8 | 4 | 4 | 1 | 4 |
| 5 | 6 | 10 | 5 | 5 | 1 | 5 |
| 6 | 7 | 12 | 6 | 6 | 1 | 6 |
| 7 | 8 | 14 | 7 | 7 | 1 | 7 |
| 8 | 9 | 16 | 8 | 8 | 1 | 8 |
| 9 | 10 | 18 | 9 | 9 | 1 | 9 |
| 10 | 11 | 20 | 10 | 10 | 1 | 10 |
| Transpiled quantum circuit | | | | | | |
| 4 | 12 | 131 | 4 | 23 | 1 | 127 |
| 5 | 14 | 132 | 5 | 29 | 1 | 127 |
| 6 | 20 | 133 | 6 | 44 | 1 | 127 |
| 7 | 21 | 134 | 7 | 48 | 1 | 127 |
| 8 | 22 | 135 | 8 | 51 | 1 | 127 |
| 9 | 25 | 136 | 9 | 60 | 1 | 127 |
| 10 | 31 | 137 | 10 | 73 | 1 | 127 |

*Table 5: W state quantum and transpiled circuit details*

| **circuit** | **depth** | **width** | **initialize** | **measu rement s** | | **no. of gates** |
|---|---|---|---|---|---|---|
| **Quantum circuit** | | | | | | |
| 4 | 2 | 8 | 1 | 4 | | 4 |
| 5 | 2 | 10 | 1 | 5 | | 5 |
| 6 | 2 | 12 | 1 | 6 | | 6 |
| 7 | 2 | 14 | 1 | 7 | | 7 |
| 8 | 2 | 16 | 1 | 8 | | 8 |
| 9 | 2 | 18 | 1 | 9 | | 9 |
| 10 | 2 | 20 | 1 | 10 | | 10 |
| **Transpiled quantum circuit** | | | | | | |
| 4 | 4 | 65 | 31 | 4 | | 99 |
| 5 | 5 | 155 | 132 | 5 | | 278 |
| 6 | 6 | 349 | 133 | 6 | | 323 |
| 7 | 7 | 737 | 134 | 7 | | 749 |
| 8 | 8 | 1826 | 135 | 8 | | 3446 |
| 9 | 9 | 3455 | 136 | 9 | | 7162 |
| 10 | 10 | 7083 | 137 | 10 | | 13626 |

The complexity of the base quantum circuits and their transpiled counterparts implemented on superconducting QPU backends is summarized in Tables 3–5. Open quantum hardware, such as the 127-qubit processors with limited execution time per user, is accessible through the IBM Quantum ecosystem. Transpilation frequently produces an effective circuit width that exceeds the nominal 127-qubit limit, even when the base circuits are designed within it. This indicates the introduction of auxiliary (logical) qubits to meet routing, connectivity, and hardware constraints.

Post-transpilation, there is a notable increase in circuit depth and gate count, which is largely driven by the underlying circuit's intrinsic complexity. The circuit depth and gate count for 10-qubit systems increase to 7083 and 13,626, respectively, as shown in Table 5, indicating the significant overhead added during hardware mapping. This amplification highlights the practical difficulties of scaling quantum circuits for real-world applications, underscoring the challenging transition from idealized circuits to hardware-executable implementations.

From this perspective, the observed variations in transpiled circuit complexity across circuit classes—such as the W state, Greenberger–Horne–Zeilinger state, and Quantum Fourier Transform—can be explained by differences in their sensitivities to hardware noise. While physically simpler circuits, such as GHZ states, show relatively lower susceptibility, circuits with larger gate counts and depth, like W-state implementations, effectively magnify the effects of material-limited decoherence. Transpiled circuit performance may serve as an indirect test of defect-driven noise in superconducting quantum devices, as this provides a direct connection between circuit-level behavior and underlying material quality.

Because of its organized decomposition and versatility, the Quantum Fourier Transform exhibits rather favorable properties for building application-oriented quantum workflows among the circuit classes examined. Nevertheless, the issue domain ultimately determines whether a circuit is appropriate, and larger-scale applications would require access to higher-qubit hardware platforms with better coherence and connectivity.

Although simulators provide a reliable baseline for optimal circuit behavior, their usefulness quickly diminishes as the number of qubits increases due to the exponential memory requirements. Real quantum hardware, on the other hand, enables scaling, but it requires the use of quantum error mitigation techniques to address noise. Additionally, effective data encoding strategies are necessary for the practical implementation of quantum circuits, especially in data-intensive applications like quantum machine learning, where the trade-off between computational resources and execution time becomes crucial [57], [58].

Thus, transpiled circuit complexity not only reflects architectural constraints but also encodes signatures of underlying material-induced decoherence in superconducting quantum processors. These observations reinforce the need for hardware-aware circuit design, where algorithmic efficiency must be co-optimized with architectural constraints to enable scalable quantum computation.

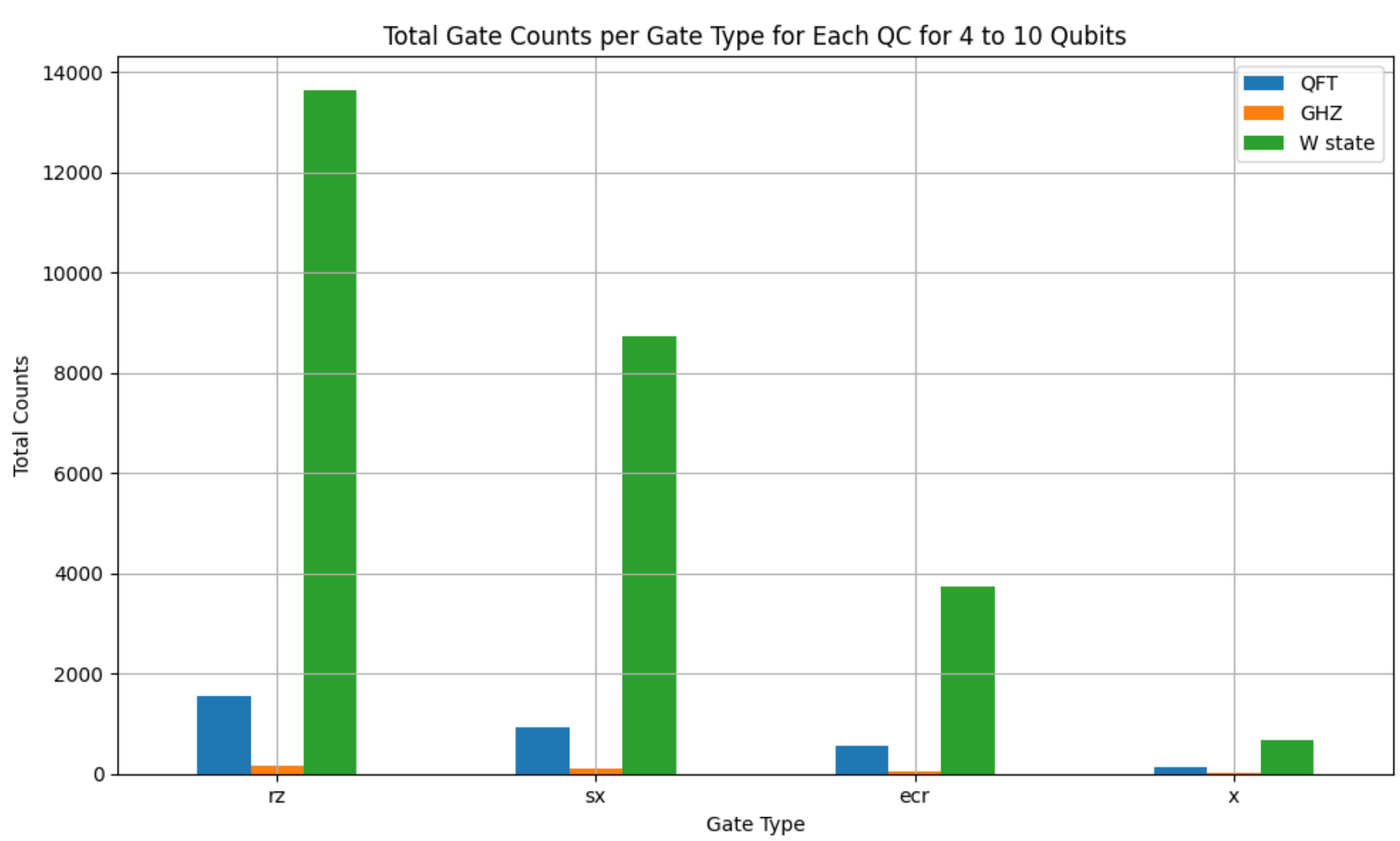

*Fig. 18: Transpiled circuits and Gate Type comparison*

Based on the breakdown in Tables III–V, Fig. 18 shows a comparative study of transpiled circuit gate distributions on superconducting physical qubits. One noteworthy finding is that single-qubit phase rotations (especially RZ gates) can reach up to 13,623 in the W-state circuits, which have a much larger gate count than both the quantum Fourier Transform and Greenberger–Horne–Zeilinger state circuits. This suggests that W-state implementations have significantly higher circuit overhead after transpilation, perhaps due to their decomposition requirements and entanglement structure.

## A. Discussion:

GHZ circuits, on the other hand, constantly exhibit the fewest gate counts, indicating relatively shallow circuit depth and structural simplicity. This is consistent with their architecture of entanglement that is more direct and involves fewer intermediate processes. Since such basic processes are frequently absorbed or substituted during gate deconstruction and optimization in hardware-aware compilation workflows, the limited existence of X gates across all circuit types is also anticipated.

The discovered gate distributions shed light on how hardware-specific calibration limitations in superconducting quantum computers interact with circuit properties, such as entanglement topology, algorithmic structure, and intended application. To guarantee compliance with the connection and native gate set of the physical device, transpilation specifically adds extra gates and changes the circuit design. To reflect the limitations imposed by actual hardware, this change not only increases gate counts but also changes gate types.

Additionally, additional logical qubits are introduced during transpilation, which is frequently necessary for routing and mapping onto limited-connectivity designs, according to analysis of circuit width and qubit allocation. As a result, the idealized circuits used in simulation settings differ significantly from the overall numbers of qubits, gates, and operations.

Crucially, the impact of noise in hardware becomes increasingly noticeable as circuit complexity rises, especially with higher qubit counts. The differences between simulator and QPU executions increase with circuit size, highlighting the cumulative impact of gate faults and decoherence, even though the simulator results remain idealized. This pattern emphasizes the usefulness of quantum error mitigation and correction techniques, particularly for circuits with greater entanglement complexity and depth.

The upper limit of the experimental analysis is set by the fact that this study is limited to a 127-qubit superconducting processor. However, the results lay the groundwork for further research on next-generation quantum processors with higher qubit counts and better coherence, enabling comparable analyses to evaluate error resilience and scalability in more sophisticated quantum systems.

Circuit integrity in superconducting quantum processors can be understood as an emergent system-level observable that reflects underlying material decoherence mechanisms. Quantum circuits are often viewed as indirect probes of material quality because of the exponential deterioration of circuit fidelity with increasing circuit depth, which suggests a direct connection to defect-induced noise sources, such as two-level systems (TLS).

# V. Conclusion

Due to their distinct structural and functional properties, foundational quantum circuits such as the Quantum Fourier Transform, Greenberger–Horne–Zeilinger state, and W state have found widespread applications across domains, including quantum communication, optimization, and emerging areas such as quantum machine learning. The required number of qubits in these applications varies significantly with problem scale and domain-specific complexity, thereby influencing circuit design and implementation strategies.

Within superconducting qubit-based quantum processing units, this work provides insights into constructing larger, more complex quantum circuits by systematically building on these fundamental circuit primitives. Such an approach enables the development of modular subsystems that can be composed to explore pathways toward achieving quantum advantage. However, as the number of qubits increases, classical simulation becomes increasingly constrained by exponential memory requirements, limiting the feasibility of large-scale simulations.

The outcomes of this study are particularly relevant for hardware-aware quantum circuit design, especially in the context of IBM Quantum systems. By analyzing circuit layouts alongside hardware constraints, the work enables better planning for compatibility with superconducting qubit architectures. This is especially critical for applications involving large datasets and high-dimensional parameter spaces, such as quantum machine learning, where circuit scalability and efficiency are essential.

Designing quantum circuits from fundamental building blocks—whether used independently or integrated with additional gates and subsystems—requires careful consideration of layout, connectivity, and optimization. In the Noisy Intermediate-Scale Quantum (NISQ) era, circuits must be adapted to conform to hardware-specific requirements before execution. In this context, transpilation is essential, transforming abstract circuits into hardware-compatible forms. A thorough understanding of its impact on circuit structure, including changes in depth, width, and gate composition, is therefore essential.

Optimizing these parameters involves inherent trade-offs between computational complexity, hardware constraints, and noise susceptibility. Consequently, effective circuit design requires balancing factors such as resource allocation, execution fidelity, and noise robustness, alongside post-processing strategies such as error mitigation. This study contributes to developing such foundational insights, enabling more informed planning and optimization of quantum circuits toward realizing practical quantum advantage.

As discussed above, a multi-layered physical architecture connecting materials to computation can be used to analyze the observed decline in circuit quality with increasing circuit depth. Through relaxation and dephasing mechanisms, tiny defects such as two-level systems (TLSs), interfacial losses, and dielectric flaws shorten coherence times on superconducting qubit platforms. At the device level, these effects, controlled by quantum decoherence, directly limit gate fidelity. Gate faults accumulate as quantum circuits become deeper and more complex, leading to a deterioration in total circuit fidelity that grows exponentially.

As a result, it is possible to think of circuit-level performance as an emergent measure of underlying material quality. In this regard, the differences between various circuit classes—such as the W-state, Greenberger–Horne–Zeilinger state, and quantum Fourier transform—reflect both the algorithmic structure and the circuits' susceptibility to hardware noise. This establishes a conceptual connection between the performance of quantum algorithms and materials engineering, suggesting that circuit benchmarking could serve as an indirect means to investigate defect-induced noise in superconducting quantum devices.

# Funding Declaration:

There is no funding for this work.